\documentclass[prd,reprint,showpacs,showkeys]{revtex4}
\usepackage{graphics}
\usepackage{eurosym}
\usepackage{amsfonts}
\usepackage{mdframed}
\usepackage{amssymb}
\usepackage{amsmath}
\usepackage{graphicx}
\usepackage[font={footnotesize,it}]{caption}

\begin{document}
\title{Tunneling of Vector Particles from Lorentzian Wormholes in 3+1
Dimensions}
\author{I. Sakalli}
\email{izzet.sakalli@emu.edu.tr}
\author{A. Ovgun}
\email{ali.ovgun@emu.edu.tr}
\affiliation{Physics Department , Eastern Mediterranean University,
Famagusta, Northern Cyprus, Mersin 10, Turkey}

\date{\today}

\begin{abstract}In this article, we consider the Hawking radiation (HR) of vector (massive
spin-1) particles from the traversable
Lorentzian  wormholes (TLWH) in 3+1 dimensions. We start by providing the
Proca equations for the  TLWH. Using the Hamilton-Jacobi (HJ)ans\"{a}tz  with the WKB approximation in the quantum tunneling method, we obtain the probabilities of the emission/absorption
modes.  Then, we derive the tunneling rate of the emitted vector particles and manage to read the standard Hawking temperature of the TLWH. The result obtained represents a negative temperature, which is also discussed.   \end{abstract}

\maketitle
\section{Introduction}

Since the influential paper of Morris and Thorne\cite{Morris}, one of the
most fascinating features of the theory of general relativity is the
potential existence of space-times with wormholes. It is believed that they
are the short-cut between otherwise distant or unconnected regions of the
universe. Topologically, wormhole space-times are the same as those of black
holes (BHs), however its throat which possesses a minimal surface maintained
in the time evolution allows a traveler to both direction. The throat is
held open by the presence of a phantom field \cite{Caldwell}. Namely,
phantom energy is precisely what is needed to support traversable wormholes.
However, this exotic quantity violates the null energy condition, which
signals the existence of the dark energy that dominates our Universe \cite%
{Caroll}.

In the mid-1970s, Stephen Hawking looked into whether BHs could radiate
thermally according to the quantum mechanics using the Wick rotation method 
\cite{Hawking,Hawking',Hawking''}. Throughout the space, short-lived
\textquotedblleft virtual\textquotedblright\ particles (pair of the real
particle and the anti-particle) continually pop into and out of the
existence. Hawking realized that if the anti-particle falls into a BH and
the real one escapes, and the BH would emit radiation, glowing like a dying
ember.\ Heuristically, there exist several derivations of the HR, such as
the Damour-Rufini method, the HJ method, and the Parikh-Wilczek tunneling
method (PWTM) (see for instance \cite%
{Hawking1,Damour,Wilczek,Wilczek1,Mann,Mann1,Christensen,Zhang1}). The
reader is referred to \cite{review} for the topical review. Meanwhile, it is
worth noting that the PWTM is only applicable to a future outer trapping
horizon of the wormhole \cite{hawyard2009}. All these methods can be used to
calculate the emission/absorbtion probabilities of the particles penetrating
the particular surface (event horizon) of the BH from the inside to the
outside of the horizon, or vice-versa, via the following relation

\begin{equation}
\Gamma =e^{-2ImS/\hslash },  \label{1}
\end{equation}

where $S$ is the action of the classically forbidden trajectory. Thus, the
Hawking temperature is derived from the tunneling rate of the emitted
particles (see for example \cite%
{Jing,Mann3,yang,Kruglov1,ran2,sharif,ran,ChenZhou,ali2,ali1}.

The remainder of this paper is organized as follows. In Sec. II, we
introduce the 3+1 dimensional TLWH \cite{kim}. Section III analyzes the
Proca equation for massive vector particles in the past outer trapping
horizon \cite{hawyard94,hayward,hayward1} geometry of the TLWH. We show that
the Proca equations amalgamated with the HJ method can be reduced to a
single equation, which makes possible to compute the probabilities of the
emission/absorption of the spin-1 particles. Then, we read the tunneling
rate of the radiated particles and use it to derive the Hawking temperature
of the TLWH. Finally, in Sec. VI, the conclusions are summarized and further
comments are added.

\section{TLWH in 3+1 Dimensions}

There is an analog of the BHs with the wormhole topology \cite{Topology}.
However, instead of the event horizon, the wormholes must have a throat,
which allows the particles to pass through it in both directions. To
construct the throat of a wormhole, an exotic matter is required. Since the
wormholes have two ends, the inside particles can naturally radiate from the
both ends.

For studying the HR of TLWH, we consider a general spherically symmetric and
dynamic wormhole with a past outer trapping horizon. As it is shown by \cite%
{kim}, this local metric can be expressed in terms of the generalized
retarded Eddington-Finkelstein coordinates as

\begin{equation}
ds^{2}=-Cdu^{2}-2dudr+r^{2}\left( d\theta ^{2}+Bd\varphi ^{2}\right) ,
\label{2}
\end{equation}

where $C=$ $1-2M/r$\ and $B=\sin ^{2}\theta $. $M$ represents the
gravitational energy in space with this symmetry,\ which is the so-called
Misner-Sharp energy \cite{MisnerSharp}. $\ $It is defined by $M=\frac{1}{2}%
r(1-\partial ^{a}r\partial _{a}r),$ which becomes $M=\frac{1}{2}r$ on a
trapping horizon. Moreover, the retarded coordinates admit that the marginal
surfaces in which $C=0$ (at horizon: $r=r_{0}$) are the past marginal
surfaces \cite{kim}.

\section{HR of Vector Particles From 3+1 Dimensional TLWH}

We start to the section by introducing the Proca equation for a curved
space-time \cite{K2,ali3}:

\begin{equation}
\frac{1}{\sqrt{-g}}\frac{\partial \left( \sqrt{-g}\psi ^{\nu \mu }\right) }{%
\partial x^{\mu }}+\frac{m^{2}}{\hbar ^{2}}\psi ^{\nu }=0,  \label{3}
\end{equation}

where the wave function for a 3+1 dimension is given by $\psi _{\nu }=(\psi
_{0},\psi _{1},\psi _{2},\psi _{3})$. Next, within the framework of the WKB
approximation, we substitute the following HJ ans\"{a}tz into Eq. (3)

\begin{equation}
\psi _{\nu }=\left( c_{0},c_{1},c_{2},c_{3}\right) e^{\frac{i}{\hbar }%
S(u,r,\theta ,\phi )},  \label{4}
\end{equation}

where $\left( c_{0},c_{1},c_{2},c_{3}\right) $ denote the arbitrary real
constants. The action $S(u,r,\theta ,\phi )$ is given by 
\begin{equation}
S(u,r,\theta ,\phi )=S_{0}(u,r,\theta ,\phi )+\hbar S_{1}(u,r,\theta ,\phi
)+\hbar ^{2}S_{2}(u,r,\theta ,\phi )+....  \label{5}
\end{equation}

Since metric (2) is symmetric, we have the Killing vectors $\partial
_{\theta }$\ and $\partial _{\phi }$.\ So, one can apply the separation of
variables method to the action $S_{0}(u,r,\theta ,\phi )$: 
\begin{equation}
S_{0}=Eu-W(r)-j\theta -k\phi ,  \label{6}
\end{equation}%
where $E$ and $(j,k)$ are energy and real angular constants, respectively.
After inserting Eqs. (4), (5), and (6) into Eq. (3), we obtain a matrix
equation $\Delta \left( c_{0,}c_{1},c_{2},c_{3}\right) ^{T}=0$ (to the
leading order in $\hbar $)$,$ which has the following non-zero the
components :

\begin{eqnarray}
\Delta _{11} &=&2B\left[ \partial _{r}W(r)\right] ^{2}r^{2},\   \notag \\
\Delta _{12} &=&\Delta _{21}=2m^{2}r^{2}B+2B\partial
_{r}W(r)Er^{2}+2Bj^{2}+2k^{2},  \notag \\
\Delta _{13} &=&-\frac{2\Delta _{31}}{r^{2}}=-2Bj\partial _{r}W(r),\   \notag
\\
\Delta _{14} &=&\frac{\Delta _{41}}{Br^{2}}=-2k\partial _{r}W(r),  \notag \\
\Delta _{22} &=&-2BCm^{2}r^{2}+2E^{2}r^{2}B-2j^{2}BC-2k^{2}C,\   \label{7n}
\\
\Delta _{23} &=&\frac{-2\Delta _{32}}{r^{2}}=2jBC\partial _{r}W(r)+2EjB, 
\notag \\
\Delta _{24} &=&\frac{\Delta _{42}}{Br^{2}}=2kC\partial _{r}W(r)+2kE,  \notag
\\
\Delta _{33} &=&m^{2}r^{2}B+2BEr^{2}\partial _{r}W(r)+r^{2}BC\left[ \partial
_{r}W(r)\right] ^{2}+k^{2},  \notag \\
\Delta _{34} &=&\frac{-\Delta _{43}}{2B}=-kj,  \notag \\
\Delta _{44} &=&-2r^{2}BC\left[ \partial _{r}W(r)\right] ^{2}-4BEr^{2}%
\partial _{r}W(r)-2B(m^{2}r^{2}+j^{2}).  \notag
\end{eqnarray}

A non-trivial solution is conditional on the termination of the determinant
of the $\Delta $-matrix ($\mbox{det}\Delta =0$). Hence, we get%
\begin{equation}
\mbox{det}\Delta =64Bm^{2}r^{2}\left\{ \frac{1}{2}r^{2}BC\left[ \partial
_{r}W(r)\right] ^{2}+BEr^{2}\partial _{r}W(r)+\frac{B}{2}\left(
m^{2}r^{2}+j^{2}\right) +\frac{k^{2}}{2}\right\} ^{3}=0.  \label{8n}
\end{equation}

Solving Eq. (8) for $W(r)$ yields 
\begin{equation}
W_{\pm }(r)=\int \left( \frac{-E}{C}\pm \sqrt{\frac{E^{2}}{C^{2}}-\frac{m^{2}%
}{C}-\frac{j^{2}}{CB^{2}r^{2}}-\frac{k^{2}}{Cr^{2}}}\right) dr.  \label{9}
\end{equation}%
In the vicinity of the horizon ($r\rightarrow r_{0}$), the above intergral
takes the following form

\begin{equation}
W_{\pm }(r)\simeq \int \left( \frac{-E}{C}\pm \frac{E}{C}\right) dr.
\label{10}
\end{equation}

According to Eq. (1), the probabilities of emitted/absorbed particles depend
on the imaginary contribution of the action. Since $C=0$ on the horizon, Eq.
(10) has a pole. Therefore, the associated contribution is obtained by
deforming the contour of integration in the upper $r$ half-plane. In short,
at the horizon, Eq. (10) becomes

\begin{equation}
W_{\pm }=i\pi \left( \frac{-E}{2\kappa |_{H}}\pm \frac{E}{2\kappa |_{H}}%
\right) .  \label{11}
\end{equation}

Whence

\begin{equation}
ImS=ImW_{\pm },  \label{12}
\end{equation}

where $\kappa |_{H}=\partial _{r}C/2$ is the surface gravity at the horizon.
It should be noted that since the throat is an outer trapping horizon, the $%
\kappa |_{H}$ is positive quantity \cite{kim,kim2}. If we set the
probability of absorption to $100\%$ (i.e., $\Gamma _{absorption}\approx
e^{-2ImW}\approx 1)$ so that we consider\ $W_{+}$ for the ingoing particles,
and consequently $W_{-}$ stands for the outgoing particles, we can compute
the tunneling rate \cite{Mann1,K2,ran2,sharif,ran} of the vector particles
as 
\begin{equation}
\Gamma =\frac{\Gamma _{emission}}{\Gamma _{absorption}}=\Gamma
_{emission}\approx e^{-2ImW_{-}}=e^{\frac{2\pi E}{\kappa |_{H}}}.  \label{13}
\end{equation}

Comparing \ Eq. (13) with the Boltzmann factor $\Gamma \approx $ $e^{-\beta
E}$ ($\beta $ is the inverse temperature), we then have

\begin{equation}
T|_{H}=-\frac{\kappa |_{H}}{2\pi },  \label{14}
\end{equation}%
where $T|_{H}$ is the Hawking temperature of the TLWH. But $T|_{H}$ is
negative, as formerly stated by \cite{kim,kim2}. The main reason of this
negativeness is the phantom energy \cite{kim,yurov}, which is located at the
throat of wormhole. Furthermore, because of the phantom energy, the ordinary
matter can travel backward in time \cite{kim2}\textbf{.}

\section{Conclusion}

In summary, we have calculated the HR of the massive vector particles from
the TLWH in 3+1 dimensions. To this end, we have used the Proca equation.
The probabilities of the vector particles crossing the trapped horizon of
the TLWH\ have been obtained by applying the HJ method. The tunneling rate
of the vector particles has been obtained, and comparing it with the
Boltzmann factor the Hawking temperature of the TLWH has been obtained.
Although the computed temperature is negative, our result is consistent with
the results of \cite{kim,kim2}. Remarkably, we infer from the the negative $%
T|_{H}$ that past outer trapping horizon of the TLWH radiate thermal phantom
energy. On the other hand, it is a fact that phantom energy radiation must
decrease both the size of the throat of the wormhole\ and its entropy.
However, this does not constitute a problem. Because the total entropy of
universe always increases, and consequently it prevents the violation of the
second law of thermodynamics \cite{Diaz}.

\end{document}